\documentclass[epsf,12pt]{article}
\usepackage{graphicx}
\begin{document}
\baselineskip=18pt
\addtolength{\arraycolsep}{-\arraycolsep}
\def\be{\begin{equation}}
\def\ee{\end{equation}}
\def\bearst{\begin{eqnarray*}}
\def\eearst{\end{eqnarray*}}
\def\peleven{\parbox{11cm}}
\def\peffec{\peight{\bearst\eearst}\hfill\peleven}
\def\pspace{\peight{\bearst\eearst}\hfill}
\def\ptwelve{\parbox{12cm}}
\def\peight{\parbox{8mm}}
\def\bear{\begin{eqnarray}}
\def\eear{\end{eqnarray}}
\def\E{{\rm e}}
\input epsf.tex
\def\tr{ \mathop{\rm tr}}
\def\atanh{\mathop{\rm atanh}}
\def\Tr{\mathop{\rm Tr}}
\def\dal{\Box} 
\def\Natural{\hbox{\hskip 1.5pt\hbox to 0pt{\hskip -2pt I\hss}N}}
\def\Integer{\>\hbox{{\sf Z}} \hskip -0.82em \hbox{{\sf Z}}\,}
\def\Rational{\hbox{\hbox to 0pt{\hskip 2.7pt \vrule height 6.5pt
                                  depth -0.2pt width 0.8pt \hss}Q}}
\def\Real{\hbox{\hskip 1.5pt\hbox to 0pt{\hskip -2pt I\hss}R}}
\def\Complex{\hbox{\hbox to 0pt{\hskip 2.7pt \vrule height 6.5pt
                                  depth -0.2pt width 0.8pt \hss}C}}
\def \ln {{\rm ln}\, }
\def \cotg {\rm cotg }
\begin{center}
{\Large\bf Numerical study of the decay amplitudes in two dimensional QCD }
\vskip 1.1cm
{\bf Elcio Abdalla$^a$ and Nelson A. Alves$^b$}\\
\vskip 0.3cm
{$^a$\it Departamento de F\'\i sica Matem\'atica, Instituto de 
F\'\i sica - USP \\
C.P. 66318, S\~ao Paulo, SP, Brazil,\\
\vskip 0.3cm
$^b$ \it Departamento de F\'\i sica e Matem\'atica, FFCLRP - USP \\
         Av. Bandeirantes 3900, CEP 014040-901 Ribeir\~ao Preto, SP, Brazil
}

\end{center}
\abstract 

After presenting a survey of theoretical results concerning the 
structure of two-dimensional QCD, we present a numerical study 
related to the mass eigenstates and the decay amplitudes of higher 
mesonic states. We discuss in detail the fate of important dynamical
points such as stability of the spectrum and the problem of
screening versus confinement in this context. We point out
differences in the large distance behaviour of the potential,
which can be responsible for the question of stability of the
spectrum, as well as whether it is finite.

\vfill\eject

\section{Introduction}
\indent

Unlike the Schwinger model\cite{schw}, Quantum Chromodynamics of massless
fermions in two dimensions is not exactly solvable\cite{aar}.
It nevertheless serves as a very useful laboratory for studying problems 
such as the bound-state spectrum and algebraic structure. These problems are 
important tools for general understanding of  realistic quantum field 
theories and are expected to realize the important features of four
dimensional Quantum Chromodynamics. In particular, exact properties may 
be derived, once one arrives at an equivalent bosonic formulation in the 
form of a gauged Wess-Zumino-Witten (WZW) action\cite{50,wittencmp}.

The first attempt to obtain the particle spectrum dates back to 1974, and was
based on the $1/N$ expansion\cite{thooft,thooft2}, where $N$ is the number 
of colours.
In this limit one is led to a bound state spectrum corresponding asymptotically
to a linearly rising Regge trajectory. The use of the principal-value 
prescription in dealing with the infrared divergencies is however 
highly ambiguous due to the non-commutative nature of principal-value 
integrals. Moreover, the result for the fermion propagator is 
tachyonic for a small fermion mass as compared with the coupling constant,
hence, in particular, for zero fermion mass. This has made 't Hooft's 
solution a controversial issue\cite{review}.

In the large $N$ approximation the gluons remain massless, since fermion 
loops do not contribute to the Feynman amplitudes. This is unlike the 
$U(1)$ case, where the 
photon acquires a mass via an intrinsic Higgs mechanism. This has led to 
the speculations that $QCD_2$ may in fact exist in two phases associated with
the weak and strong coupling regimes. In this picture, the large $N$ limit
would correspond to the weak-coupling limit ('t Hooft's phase), with
massless gluons and a mesonic spectrum described by a Regge trajectory. 
In such a case, the Regge behaviour of the mesonic spectrum is compatible
with confinement. In the strong coupling regime (Higgs
phase), on the other hand, the gluons would be massive, and the original
$SU(N)_c$-symmetry  would be broken down to the maximal abelian subgroup
(torus) of $SU(N)_c$.

The behaviour of the theory in the strong or weak coupling limits is
rather subtle. The theory is asymptotically free. In the 
strong coupling limit, it is expected to be in the confining phase:
in the infinite infrared cut-off limit  quarks disappear from the spectrum, 
which consists of mesons lying approximately on a Regge trajectory.

The problem of screening and confinement can
however be analysed along the lines of the $U(1)$ case. However, unlike the
$U(1)$ case, the screening phase prevails in the non-abelian 
theory\cite{gross-kle-etc,amz}.

\subsection{\it QCD$_2$ in the local decoupled formulation and BRST 
constraints}

The partition function of two-dimensional QCD in the
fermionic formulation (before gauge fixing) is given by the expression
\be
{\cal Z}=\int{\cal D} A_+{\cal D}A_-\int{\cal D}\psi{\cal D}\bar\psi
e^{iS[A,\psi,\bar\psi]}\label{11qcd2partitionfunction}
\ee
with the action
\be
{\cal S}[\!A\!,\psi\!,\!\bar\psi\!]=
\int\! d^2x\!\left[-\frac{1}{4}\!\tr F^{^{\mu\nu}}
F_{_{\mu\nu}}+\psi^\dagger_1(i\partial_++eA_+)\psi_1+\psi^\dagger_2(i\partial
_-+eA_-)\psi_2\right].\label{qcd2classaction}
\ee
We can obtain a bosonic formulation of the theory, in such a way
that structural relations,  hidden in the fermionic formulation are made
clearer in the bosonic counterpart.
We first make some useful change of variables, obtaining a formulation
in terms of matrix-valued fields which decouple at the partition function
level, but which are not totally decoupled, due to the gauge symmetries
of the theory.

We can also arrive at a factorized form of the partition function 
(\ref{11qcd2partitionfunction}) by pa\-ra\-me\-trizing $A_\pm$ as
\be\label{aparametrization}
eA_+=U^{-1}i\partial_+U\quad ,\quad A_-=Vi\partial_- V^{-1}
\ee
 as well as performing a chiral rotation,
\be 
\psi_1\to\psi_1^{(0)}\equiv U\psi_1,\quad \psi_2\to\psi_2^{(0)}
=V^{-1}\psi_2\quad .\label{ch11transf}
\ee
We arrive at
\be
{\cal Z}={\cal Z}_F^{(0)}\int{\cal D}U{\cal D}W{\cal J}_G[W]{\cal J}_F
[W]e^{i {\cal S}_{YM}[W]}\quad ,\label{ch11bosonicpf}
\ee
where  ${\cal Z}^{(0)}_F$ is the partition function of free fermions,
and ${\cal S}_{YM}$ is the Yang-Mills action given by
\be
{\cal S}_{YM}[W] = \int d^2x \lbrack E^2+E\partial_+ 
(Wi\partial_- W^{-1})\rbrack \label{ch11ymaction1}
\ee
with $W=UV$.

The field strength tensor $F_{01}$ is given in terms of $W$ and $U$ or $V$
in either of the two alternative forms
\be
F_{01}=-\frac{1}{2}U^{-1}[\partial_+(Wi\partial_- W^{-1})] U=
\frac{1}{2}V[\partial_-(W^{-1}i\partial_+ W)] V^{-1}\quad . 
\label{ch11stresstensor2}
\ee

The Jacobian ${\cal J}_F$ is given, following Fujikawa, by
\be
{\cal J}_F[UV] =e^{-i\Gamma[UV]} \label{jacobian}
\ee
while the determinant of the adjoint Dirac operator is
\be
{\cal J}_G[UV]=e^{-ic_V\Gamma[UV]}(\det i\partial_+)_{adj}
(\det i\partial_-)_{adj}\quad ,\label{ch11jacobiadj} 
\ee
where $c_V$ is the second Casimir of the group in question with
the normalization $f_{acd}f_{bcd}=c_V\delta_{ab}$
of the structure constants, and $\Gamma [UV]$ is the Wess-Zumino-Witten 
action. Representing $(\det i\partial_\pm)_{adj}$
in terms of ghosts and choosing the gauge $U=1$, we obtain
\be
{\cal Z}={\cal Z}^{(0)}_F {\cal Z}^{(0)}_{gh}{\cal Z}
_W\quad ,\label{ch11localpf} 
\ee
where ${\cal Z}^{(0)}_{gh}={\cal Z}^{(0)}_{gh+}{\cal Z}^{(0)}_{gh-}$
and
\bear
{\cal Z}_W&=&\int{\cal D}W e^{-i(1+c_V)\Gamma[W]}
e^{i{\cal S}_{YM}[W]}=\int{\cal D}W e^{iS_{eff}[W]}\quad ,
\label{ch11localwpf}\\ 
\eear
where we introduced the $W$--effective action
\bear
S_{eff}[W] &=& S_{YM} -(c_V +1)\Gamma [W]\quad . \label{localeffecaction} 
\eear
The effective action has two terms, one corresponding to a 
WZW action with a 
negative coefficient, and the Yang-Mills action written in the 
form (\ref{ch11ymaction1}).

We refer to (\ref{ch11localpf}) as the ``local decoupled''
partition function. As seen from (\ref{ch11localwpf}), the ghosts
$b_\pm^{(0)}$ are canonically conjugate to $c_\pm^{(0)}$ and have 
Grassmann parity $+1$. We assign to them the ghost number 
$gh\# = -1$ and $gh\# = +1$, respectively.

The dimensionality of the direct-product space
${\cal H}_F^{(0)}\otimes {\cal H}_{gh}^{(0)}\otimes {\cal H}_W$
associated with the partition function (\ref{ch11localpf}) is
larger than that of the physical Hilbert space of the original
fermionic formulation. Hence there must exist constrains imposing restrictions
on the representations which are allowed in ${\cal H}_{phys}$. In order
to discover these constraints we observe that the partition function
is separately invariant under the following nilpotent transformations
\cite{cr,crs}\\
\parbox{5cm}{\bearst
&&W\delta W^{-1}=- c_-^{(0)}\; , \\
&&\delta\psi^{(0)}_1= c_-^{(0)}\psi_1^{(0)}\; ,\;
\delta\psi^{(0)}_2=0\; ,\\
&&\delta c_-^{(0)}=\frac{1}{2}\{c_-^{(0)} , c_-
^{(0)}\} \; , \;\delta c_+^{(0)}=0\; ,\\
&&\delta b^{(0)}_- =\Omega_-
\; , \;
\delta b_+^{(0)}=0\quad ,\eearst}
\hfill
\parbox{5cm}{\bearst
&&W^{-1}\delta W=- c_+^{(0)}\; ,\\
&&\delta\psi_1^{(0)}=0\; ,\; \delta\psi_2^{(0)}=
c_+^{(0)}\psi_2^{(0)}\; ,\\
&&\delta c^{(0)}_-=0\; ,\; \delta c_+^{(0)}=\frac{1}{2}
\left\{c_+^{(0)},c_+^{(0)}\right\}\; ,\\
&&\delta b_-^{(0)}=0\; ,\; \delta b_+^{(0)}=\Omega_+\; ,
\eearst}\hfill\peight{\bear\label{localbrstplus}\eear}\\ 
where $\delta$ denotes the variation graded with respect to Grassmann
parity, and 
$\Omega_\mp$ are given by\\
\parbox{10cm}{\bearst
\Omega_- &=& -\frac{1}{4e^2}{\cal D}_-(W)(\partial_+(Wi\partial_-
W^{-1}))-\left({1+c_V}\right)
J_- (W) +j_- \\
\Omega_+&=&-\frac{1}{4e^2}{{\cal D}}_+(W)(\partial_-(W^{-1}i\partial_+
W))-\left(1+c_V\right)
J_+ (W)+j_+
\eearst}\hfill\peight{\bear\label{omegaminusplus}\eear}\\ 
with
\bear
&&J_- (W)= {1\over 4\pi}Wi\partial_- W^{-1}\, ,\quad
J_+ (W) = {1\over 4\pi} W^{-1}i\partial_+W\; ,\label{jwzwpm}\\ 
&&j_-=\psi_1^{(0)}\psi_1^{(0)\dagger}+\{b_-^{(0)},c_-^{(0)}\}
\, , \quad
j_+=\psi_2^{(0)}\psi_2^{(0)\dagger}+\{b_+^{(0)},c_+^{(0)}\}\; . 
\label{jzeropm}\eear
These transformations are easily derived by departing from
the Yang-Mills action\cite{cr,crs}.

The corresponding BRST currents,
as obtained via the usual Noether construction, are found to be
\be
J_\mp  =\tr c^{(0)}_-\left[\Omega_\mp 
-\frac{1}{2}\{b_\mp^{(0)},c_\mp^{(0)}\}\right]\quad , 
\label{noetherbrstmp} 
\ee
with $\partial_+J_- =0$, and $\partial_-J_+ =0$.

Remarkably enough, the nilpotent symmetries lead to  currents $J_-$ and
$J_+$ which only depend on the variable $x^-$ and $x^+$, respectively.

The on-shell nilpotency of the corresponding conserved charges
\be
Q_\pm =\int dx^1 J_\pm (x^\pm)\label{correspconscharges} 
\ee
follows from the first-class character of the operators $\Omega^a_\pm
=\tr \left( t^a\Omega_\pm\right)$.
\subsection{\it $QCD_2$ in the non-local decoupled formulation and BRST 
constraints}

The partition function represented by the standard expression
(\ref{ch11localwpf}) contains fields which are mixtures of massive and 
massless modes, of positive and negative norms respectively, coupled by the 
constraints. In the following we dissociate these degrees of freedom by a 
suitable transformation. We shall thereby be lead to an alternative
nonlocal representation of the partition function, useful for learning 
certain structural properties.

Following Ref.  \cite{aaijmpa}, we  make in (\ref{ch11ymaction1}) the 
change of variables $E\to\beta$ defined by
\be
\partial_+E = \left(\frac{1+c_V}{2\pi}\right)\beta^{-1}i\partial_+\beta\quad .
\label{eofbeta} 
\ee
The Jacobian  associated with this change of variables
is
\be
{{\cal D}}E =  
\det i{\cal D}_+ (\beta) {{\cal D}}\beta\label{deofdbeta} 
\ee
where we have suppressed the constant $\det \partial_+$ which will not play
any role in the discussion to follow. Making use of the determinant of the
fermionic operator in the fundamental or in the adjoint representation  and 
representing $\left( \det i\partial_+\right)$ as a functional integral over 
ghost fields $\hat b_-$ and $\hat c_-$, we have, after decoupling the ghosts,
\bear
{\cal Z}&=&{\cal Z} _F^{(0)}{\cal Z}_{gh}^{(0)} \hat{\cal  Z}_{gh-}^{(0)}
\int{\cal D}W\int{\cal D}
\beta \exp\{-i(1+c_V)[\Gamma[W]+\Gamma[\beta]
\nonumber\\
& & -\frac{1}{4\pi}\int tr(\beta^{-1}
\partial_+\beta W\partial_-W^{-1})]\} \label{ch11nonlocalpf}\\ 
& & \times \exp({ i\Gamma[\beta]})
\exp\left\{{i\left(\frac{1+c_V}{2\pi}\right)^2
e^2\int\frac{1}{2}tr\left[\partial^{-1}_+(\beta^{-1}\partial_+\beta
\right]^2}\right\} \quad , \nonumber
\eear
where 
\be
\hat{\cal  Z}_{gh-}^{(0)} = \int{\cal D}\hat b^{(0)}_-{\cal D}\hat 
c_-^{(0)} e^{i\int d^2 x\tr\hat 
b_-^{(0)}i\partial_+\hat c^{(0)}_-}\quad .\label{ghosthatminuspf} 
\ee
Using the Polyakov-Wiegmann identity, \cite{aar}
and making the change of variable $W\to\beta W=\tilde W$,
we are left with
\be
{\cal Z}={\cal Z}^{(0)}_F{\cal Z}^{(0)}_{gh} \hat{\cal  Z}_{gh-}^{(0)}
{\cal Z}_{\tilde W}{\cal Z}_\beta\quad , \label{z=prodzsnonlocal} 
\ee
where
\be
{\cal Z}_\beta=\int{{\cal D}}\beta \exp\left\{{i\Gamma[\beta]
+i\left(\frac{1+c_V}{2\pi}
\right)^2e^2\int \frac{1}{2}tr\left[\partial_+^{-1}
(\beta^{-1}\partial_+\beta)\right]^2}\right\}\label{zbeta} 
\ee
and
\be
{\cal Z}_{\tilde W}=\int{\cal D}\tilde W \exp[{-i(1+c_V)\Gamma[\tilde W]}]
\label{negativemetricfieldpf} 
\ee
is the partition function of a WZW field of level $-(1+c_V)$.

\subsection{\it Massive two-dimensional QCD}

The BRST symmetries of the physical states in massless QCD$_2$ are also the 
symmetries which should be imposed on the physical states in the massive case.
For massive fermions the functional determinant of the Dirac operator, an 
essential ingredient for arriving at the bosonised form of the QCD$_2$ 
partition function, can no longer be computed in closed form, and one must 
resort to the so-called adiabatic principle of form invariance\cite{aar}. 
Equivalently, one can start with a perturbative expansion in powers of the 
mass, as given by
\be
\sum {1\over n!}M^n\left\lbrack\int{\rm d}^2 x
\overline\psi\psi\right\rbrack ^n\quad . 
\ee
Afterwards, we use the (massless) bosonization formulae and re-exponentiate 
the result. In this approach, the mass term is given in terms of a 
bosonic field $g_\psi $ of the massless theory by\cite{gepner}
\bearst
S_m =- M\int \overline \psi\psi = M\mu\int \tr (g_\psi + g_\psi^{-1})\quad ,
\eearst
where $\mu$ is an arbitrary massive parameter whose value depends  on the
renormalization  prescription for the mass operator.\cite{aar}

Defining $m^2 = M\mu$, we re-exponentiate the mass term. Going through the 
changes of variables leading to (\ref{z=prodzsnonlocal}), one arrives at the 
following alternative forms for the mass term when expressed in terms of the 
fields of the non-local formulation
\be
S_m = m^2 \int\tr (g\tilde\Sigma^{-1} \beta + \beta^{-1} \tilde
\Sigma g^{-1})\quad . 
\label{2.3}
\ee

The corresponding effective action of massive QCD$_2$ in the non-local 
formulation reads
\be
S=S_{YM}\lbrack \beta, B\rbrack +S_m\lbrack g,\beta,\tilde\Sigma\rbrack +
\Gamma\lbrack g\rbrack  + \Gamma \lbrack \beta\rbrack  
- (c_V+1)\Gamma\lbrack \tilde\Sigma\rbrack +S_{gh}+
 \hat S_{gh-}\quad . \label{massive-boson-action} 
\ee

We thus see that the associated partition function no longer 
factorizes. Nevertheless, there still exist BRST currents which are either 
right- or left-moving, just as in the massless case.

The action (\ref{massive-boson-action}) exhibits various symmetries of the 
BRST type;  however, not all of them lead to nilpotent charges. The variations
are graded with respect to Grassmann number. The equations of motion obtained 
from action (\ref{massive-boson-action}) read
\bear
{1\over 4\pi}\partial_+(g\partial_-g^{-1}) = &&  m^2 (g\tilde
\Sigma^{-1}\beta -
\beta^{-1} \tilde\Sigma g^{-1})\; ,\label{massiveeqg}\\ 
-{c_V+1\over 4\pi}\partial_+(\tilde\Sigma \partial_-\tilde\Sigma^{-1}) =
\, && m^2 (\tilde\Sigma g^
{-1}\beta^{-1} - \beta g\tilde \Sigma^{-1})\; , 
\label{massiveeqtildesigma}\\ 
{1\over 4\pi} \partial_+ (\beta \partial_-\beta^{-1}) +
i \lambda\partial _+(\beta B
\beta^{-1}) =
\, && m^2 (\beta g \tilde\Sigma^{-1} - \tilde
\Sigma g^{-1}\beta^{-1})\; , \label{massiveeqbeta}\\ 
-{1\over 4\pi}\partial_-(\beta^{-1}\partial_+\beta)+&& i\lambda\lbrack
\beta^{-1} \partial_+ \beta, B\rbrack  +  \nonumber\\
i \lambda \partial_+B = \, && m^2 (g\tilde\Sigma^{-1}\beta -
\beta ^{-1}\tilde\Sigma g^{-1})\quad ,\label{massiveeqbeta2}\\
\partial_+^2 B = \, && \lambda (\beta ^{-1}i\partial_+\beta)\; ,
\label{massiveeqb}\\ 
\partial_\pm b_\mp = \, && 0 \quad ,
\quad \partial _\pm c_\mp =0 \quad ,\label{massiveeqgh}
\eear
with an analogous set of equations involving a so called {\it prime} 
sector\cite{cr}, and where $\lambda = \frac {c_V+1}{2\pi}e$.
Notice that the mass term can be transformed from one equation 
to another, by a suitable conjugation. Making use of Eqs. 
(\ref{massiveeqg}-\ref{massiveeqgh}), 
the Noether currents are constructed in the standard fashion: we make a 
general BRST variation of the action, without  using the equations of 
motion, and equate the result to the on-shell variation, taking into 
account terms arising from partial integrations. The only subtlety in 
this procedure concerns the WZW term, which only contributes off-shell to 
the variation. The four conserved Noether currents are found to be
\bear
J_\pm =\, && \tr \left( c_\pm\Omega_+ -
{1\over 2} b_\pm\{c_\pm, c_\pm\}\right)
\quad ,\label{2.10a}\\
\hat J_\pm=\, && \tr \left(\hat c_\pm \hat\Omega_\pm -
{1\over 2}\hat  b_\pm \{\hat c_\pm, \hat c_\pm\}\right)\quad ,\label{2.10d}
\eear
where the constraints above generally denoted as $\Omega $ are given by 
\bear
\Omega_+  &=& \left({1\over 4\pi}g^{-1}i\partial _+g -
{c_V +1\over 4\pi}
\tilde\Sigma^{-1}i\partial_+\tilde\Sigma + \{b_+, c_+\}\right)\quad,
\label{2.8a}\\
\Omega_-  &=& \left( {1\over 4\pi}g i\partial _- g^{-1} -
{c_V +1\over 4\pi}\tilde\Sigma^\prime i\partial_-\tilde\Sigma^{\prime -1} +
\{b_-, c_-\}  \right)\quad,
\label{2.8b}\\
\hat\Omega_- &=&   \left( {1\over 4\pi}\beta i\partial _- \beta^{-1} -
 {c_V+1\over 4\pi}\tilde \Sigma i\partial_-\tilde\Sigma^{-1} -
 \lambda \beta B \beta^{-1} +
\{b_-, c_-\}\right) \quad , \label{2.8c}\\
\hat\Omega_+ &=& \! \left(\! {1\over 4\pi}\!\beta^{\prime -1} i\partial _+
\beta^\prime\! -\! {c_V+1\over 4\pi}\tilde \Sigma^{\prime -1} i\partial_+
\tilde\Sigma^\prime\! -\! \lambda \beta^{\prime -1} B^\prime \beta^\prime\! +
\!\{\hat b_+,\hat c_+\}\right) \, . \label{2.8d}
\eear
(see \cite{cr,crs} for definitions 
concerning the primed sector, connected to the unprimed one by a nonlocal 
transformation).

From the current conservation laws
\be
\partial_\mp J_\pm=0 \quad ,\quad \partial _\pm\hat J_\mp =0 
\quad ,\label{2.9}
\ee
one infers that $\Omega_-$ and $\hat \Omega_-$ 
are right-moving, while $\Omega_+$ and $\hat\Omega_+$ are left-moving. 
Indeed, making use of the equations of motion 
(\ref{massiveeqg}-\ref{massiveeqgh}) 
one readily checks that the operators 
$\Omega_\pm$, $\hat \Omega_\pm$ satisfy
\be
\partial_\mp\Omega_\pm=0 \quad ,\quad \partial _\pm\hat\Omega \mp=0 
\quad ,\label{2.12}
\ee
consistent with the conservation laws (\ref{2.9}). In Ref. \cite{cr} it 
has been argued that gauge invariant bilinears are the physical states of 
the theory. There is of course the problem of whether we have to consider
or not that the physical states are annihilated by the non local constraints,
but that problem goes beyond the scope of the present paper.

\subsection{\it Screening in two-dimensional QCD}
Let us reconsider the problem of screening and confinement.
We shall concentrate on the case of single flavour QCD, and merely comment 
on the general case at the end of the section.

We proceed by first considering the case of massless fermions and compute
the inter-quark potential. We introduce a pair of classical colour 
charges of strength $q=q^at^a$ separated by a distance $L$. Such a pair
is introduced in the action (\ref{massive-boson-action}) by means of 
the substitution
\be
i(\beta^{-1}\partial_+\beta)^a\longrightarrow 
i(\beta^{-1}\partial_+\beta)^a-{2\pi\over e}q^a\bigg(\delta(x-{L\over 
2})-\delta(x+{L\over 2})\bigg)\quad ,\label{11externalchargesubstitution}
\ee
where $a$ is a definite colour index. This adds the following term to 
the action\footnote{This corresponds to minus the same term added to the 
Hamiltonian.}
\be
V(L)=\Delta S=S_q-S=-(c_V+1)
q^a\bigg(B^a(L/2)-B^a(-L/2)\bigg)\quad .\label{11potential-masslesscase} 
\ee
The equation of motion for $B^a$ is now replaced by
\be
\partial_+^2B^a=i\lambda(\beta^{-1}\partial_+\beta)^a
-(c_V+1)q^a\bigg(\delta(x-{L\over 2})-\delta(x+{L\over 2})\bigg),
\ee
which implies, upon substitution into the equation of motion for the 
$\beta$-field,
\begin{eqnarray}
& &\partial_+\left({i\over 4\pi\lambda}\partial_-\partial_+ B+[\partial_+ 
B,B]+i\lambda B\right)=\nonumber\\
& &\qquad\left({-iq\over 2e}\partial_-
+(c_V+1)[q,B]\right)
\lbrack\delta(x-{L\over 2})-\delta(x+{L\over 
2})\rbrack\quad .\label{11massivebetaeqwithextch}
\end{eqnarray}
We look for solutions of (\ref{11massivebetaeqwithextch}) with a fixed global 
orientation in colour space.\footnote{Note that this is a non-trivial
input, since we have no longer the freedom of choosing a gauge in which
such an Ansatz could be realized.}  We thus make  $B^a=q^a f(x)$. This
renders the problem abelian. We thus infer that the potential 
(\ref{11potential-masslesscase}) has the form
\be
V(L)= {(c_v+1)\sqrt\pi\over 2}{q^2\over e}(1-e^{-2\sqrt\pi\lambda L})
\label{11potential-masslesscasefin}
\ee
which implies that the system is in a screening phase.

We now turn to the case of massive fermions. Taking the external charge to lie
in the direction $t^2$ of $SU(N)$ space, our Ansatz for $B^a$ leads one to look
for solutions with $g$, $\beta$ and $\Sigma$ parametrized as
\be
g=e^{i2\sqrt\pi\varphi\sigma_2},\qquad \beta=e^{i2\sqrt\pi 
E\sigma_2},\qquad \Sigma=e^{-i2\sqrt\pi\eta\sigma_2}\quad .
\label{gbetawparametrizationt2}
\ee
The equations of motion (\ref{massiveeqg}-\ref{massiveeqb})
are replaced by a set of coupled sine-Gordon type equations. As is
well known\cite{aar}, the solution of the classical equations of motion in
the bosonic version contains quantum mechanical information from the
fermionic theory. We find, after solving them, the result\cite{amz}
\begin{eqnarray}
&&V(L)={(c_V+1)^2q^2\over 2}\times\nonumber\\
& &\left[ 
\left({4\pi\lambda^2-m_-^2\over 
m_+^2-m_-^2}\right)\!\!\!\left({1-e^{-m_+L}\over m_+}\right)
+\left({m_+^2-4\pi\lambda^2\over 
m_+^2-m_-^2}\right)\left({1-e^{-m_-L}\over 
m_-}\right)\right]\; ,\label{interquarkpotential}
\end{eqnarray}
where
\bear
\epsilon&=&{c_V\over (c_V+1)}\quad ,\quad  q_+={2\sqrt\pi(c_V+1)q\over 
(1+\epsilon a^2)}\; ,\nonumber \\
a&=&-{8\pi  m^2\over m_+^2- 16\epsilon  m^2}\quad ,\quad
q_-={2\sqrt\pi\epsilon a (c_V+1) q\over (1+\epsilon
a^2)} \label{11allparameters} \\
m_{\pm}^2&=&2\pi\lbrack\left(\lambda^2+(1+\epsilon)2  m^2\right)
\pm\sqrt{\left(\lambda^2+(1+\epsilon)2m^2\right)^2-8\epsilon\lambda^2  
m^2}\rbrack,\nonumber\\
Q_\pm&=&q_\pm \left[\Theta(x-{L\over 2})-\Theta(x+{L\over 2})\right]\; .
\eear

Thus we find two mass scales given by $m_+$ and $m_-$. Both these 
scales correspond to screening-type contributions if $c_V\not =0$.

Next, we compare the above results for the potential with those 
obtained for the Schwinger model. In the abelian case, 
the combination of the matter boson $\varphi$ and the negative metric scalar 
$\eta$ gives rise to the $\theta$-angle. That is, the combination 
$\Phi\equiv\varphi+\eta=\theta $ 
appears in the mass term. When fermions are massless, the electric field
and the matter boson decouple. However, due to a Higgs mechanism, the 
electric field acquires a mass and, therefore, a long-range force does 
not exist. This leads to a pure screening potential. On the other hand,
for massive fermions, the electric field couples to the matter boson $\Phi$.
Yet, $\Phi=\psi_{c_V=0}$, and hence, it remains massless. The coupling
to $\Phi$ via the mass term is the origin of the long-range force (linearly
rising potential) in the massive U(1) case, where the potential 
is confining. 
On the other hand, the expression (\ref{interquarkpotential}) for the 
potential indicates the absence of a long-range force in the non-abelian 
case. 

The abelian potential can also be obtained from (\ref{interquarkpotential})
by taking the limit $c_V\to 0$. In this limit, the mass scale 
$m_-$ tends to zero and we recover the linearly rising potential, 
signaling confinement.

\subsection{\it Equations of motion and higher conservation equations}

We are now going to deal with the action $S_{eff}$ given in 
(\ref{ch11localwpf}), obtaining further important information.

Due to the presence of higher derivatives in that action, it is
convenient to introduce an auxiliary field and rewrite it in the equivalent 
form (\ref{ch11ymaction1}).

The equation of motion of the $W$-field is easily computed. The WZW 
contribution has been obtained in \cite{ar-prd}, and the Yang-Mills action
leads to an extra term. We obtain
\bear
&&\lbrack {c_V+1\over 4\pi}\partial_+ +\frac {\partial_+\partial_-}
{(4\pi\mu)^2}\rbrack \left( W\partial_-W^{-1}\right)
-\frac {\partial_+} {(4\pi\mu)^2}\lbrack W\partial_-W^{-1},
\partial_+\left( W\partial_-W^{-1}\right)\rbrack\equiv\nonumber\\
&&\lbrack {c_V+1\over 4\pi}\partial_+ +{1\over (4\pi\mu)^2}\partial_+
{\cal D}_-\rbrack \left( W\partial_-W^{-1}\right)=0 .
\label{eqmotionw}
\eear

We can  list the relevant field operators appearing in the definition of the
conservation law (\ref{eqmotionw}), that is
\be
I^W_-= {1\over 4\pi} (c_V +1) J^W_- + {1\over (4\pi\mu)^2} \partial_+ 
\partial_-J^W_- - {1\over (4\pi\mu)^2} [ J^W_-, \partial_+J^W_-]\quad ,
\label{conslaww}
\ee
with $\partial _+  I^W_- = 0$, and $J^W_- = W\partial _-W^{-1}$.
It is straightforward to compute the Poisson algebra, using the canonical
formalism, which in the bosonic formulation includes quantum corrections. 
We have
\bear
\left\{I^W_{ij}(t,x), I^W_{kl}(t,y) \right\}  &=&  I^W_{\lbrack kj
 }
\delta_{ il\rbrack } \delta(x^1\!-\!y^1)-\alpha\delta^{il}
\delta^{kj}\delta'(x^1-y^1)\quad,\nonumber\\
\left\{ I^W_{ij}(t,x), J^W_{-kl}(t,y) \right\}&=& J^W_{-\lbrack 
kj }\delta_{ il\rbrack } \delta(x^1-y^1)
+ 2\delta_{il}\delta_{kj}\delta'(x^1-y^1)\quad ,\nonumber\\ 
\left\{ J^W_{ij}(t,x), J^W_{-kl}(t,y^1) \right\}&=& 0\quad ,
\label{fullpoissonconslaw}
\eear
where $\alpha = {1\over 2\pi}(c_V+1)$, and the indices in the right hand side
have been appropriately antisimetrized, as denoted by the bracketts
in the indices of the current $I$ and of the Kronecker delta $\delta_{ij}$.
We thus obtain a current algebra for $I_-^W$, acting on $J^W_-$ with a central 
extension.

\subsection{\it Dual case-non local formulation}

At the Lagrangian level, we find the Euler-Lagrange equations for
$\beta$ from the perturbed WZW action (\ref{zbeta}), that is,
\bear
\delta \Gamma[\beta ]&=& \left[{1\over 4\pi}\partial_-(\beta^{-1}\partial_+ 
\beta)\right]\beta^{-1}\delta\beta\quad ,\\ 
\delta\Delta(\beta) & =& 2\Big( \partial_+^{-1} 
(\beta^{-1}\partial_+\beta)  - \big[ \partial_+^{-2} 
(\beta^{-1}\partial_+\beta),(\beta^{-1}\partial_+\beta)
\big] \Big)\beta^{-1}\delta \beta\quad .
\eear

We define the current components
\bear
J_+^\beta&=& \beta^{-1}\partial_+ \beta\quad ,\\ 
J_-^\beta&= &-4\pi\mu^2\partial_+^{-2}J_+^\beta=-4\pi\mu^{2} 
\partial_+^{-2}(\beta^{-1}\partial_+\beta)\quad ,\label{currentbeta}
\eear
which summarize the $\beta $ equation of motion as a zero-curvature condition
given by
\be 
[{\cal L},{\cal L}]=[\partial_++J_+^\beta,\partial_- +J_-^\beta]=\partial_- 
J_+^\beta-\partial_+J_-^\beta+[J_-^\beta, J_+^\beta ] =0 \quad .
\label{laxpaircandidate}
\ee 
This is not a Lax pair, as {\it e.g.} in the usual non-linear $\sigma$-models, 
where $J^\beta_\mu$ is a conserved current and a conserved 
non-local charge is obtained. However, to a certain extent, the 
situation is simpler in the present case, due to the rather unusual form of 
the currents. This permits us to write the commutator
as a total derivative, in such a way that in terms of the current 
$J_-^\beta$ we have
\be 
\partial_+\left(4\pi\mu^2J_-^\beta+\partial_+\partial_-J_-^\beta+[J_-^\beta,
\partial_+J_-^\beta]\right)=0\quad .\label{consvlawbeta}
\ee

Therefore the quantity
\be 
I_-^\beta (x^-) = 4\pi \mu^2 J_-^\beta (x^+,x^-) + \partial_+ \partial_-
J_-^\beta (x^+,x^-) + [J_-^\beta (x^+,x^-), \partial_+ J_-^\beta(x^+,x^-)]
 \quad \label{iconservedbeta}
\ee
does not depend on $x^+$, and it is a simple matter to derive an infinite 
number of conservation laws from the above.

Canonical quantization proceeds straightforwardly, and as a consequence we can
compute the algebra of conserved currents, which is analogous to
(\ref{fullpoissonconslaw}).

We are thus led to speculate whether two-dimensional QCD contains an 
integrable system\cite{aar}. The theory corresponds to an 
off-critical perturbation of the WZW action. If we write $\beta=
\E^{i\phi}\sim 1+i\phi$,  we verify that the perturbing term 
corresponds to a mass term for $\phi$. The next natural step is to obtain 
the algebra obeyed by (\ref{iconservedbeta}), and its representation. 
However, there is a difficulty presented by the non-locality of the 
perturbation.

\subsection{\it Algebraic aspects of QCD$_2$ and integrability}

We saw that two-dimensional QCD, although not exactly soluble, in 
terms of free fields, is a theory from which some valuable results may be 
obtained. The $1/N$ expansion reveals a simple spectrum valid for weak 
coupling, while the strong coupling offers the possibility of understanding 
the baryon as a generalized sine-Gordon soliton. 

All such results point to a relatively simple structure, which could be
mirrored by an underlying spectrum generating algebra. In fact such an
algebra does exist. In the above-mentioned case of the large-$N$ expansion 
of pure QCD$_2$, one finds a $W_\infty$ spectrum generating algebra related 
to area-preserving diffeomorphisms of the Nambu-Goto action. For 
gauge-invariant bilinears in the Fermi fields this algebra can be 
constructed \cite{dmw}. It also appears in the description of the quantum 
Hall effect. Moreover, pure QCD$_2$ is equivalent to the $c=1$ matrix 
model, which also has a representation in terms of non-relativistic fermions.
The problem is also related to the Calogero-Sutherland models.  The mass 
eigenstates build a representation of the $W_\infty$ algebra as found 
in \cite{dmw}.

In spite of the hints toward a possible integrable structure in 
two-dim\-en\-sion\-al QCD the problem still remains largely 
open\cite{frishman}.
A possible explanation was also given in \cite{abdmoha}. The problem
points to a better understanding of the mesonic spectrum of the theory.
't Hooft's results implying an infinite number of mesons obeying Regge
behaviour is compatible with confinement. However, results obtained by several
authors \cite{gross-kle-etc,amz,frish-sonn} point to a screening phase.
In that case the quark anti-quark potential looks too shallow to support
an infinite number of bound states.

Here we investigate the decay amplitudes of two-dimensional QCD using
a refined numerical analysis, which will permit to account for the
details of the dynamics.

\section{Numerical backup}

We have investigated the behaviour of the decay amplitudes of 't Hooft
mesons in the theory, in the large $N$ limit. 

One of the strong reasons to study decay amplitudes is to learn about the
stability of the 't Hooft mesons, namely answer whether they might represent
soliton solutions of two-dimensional QCD. Moreover, these mesons are a probe
into the long range force: indeed, if two-dimensional QCD is confining
there is a linearly rising potential which acomodates an infinite number of
bound states. On the other hand, screening would imply a shallow potential,
flatening at infinity, implying a finite number of mesons.

\begin{figure}[!ht]
\begin{center}
\begin{minipage}[t]{0.85\textwidth}
\centering
\includegraphics[width=3.2in,angle=-90]{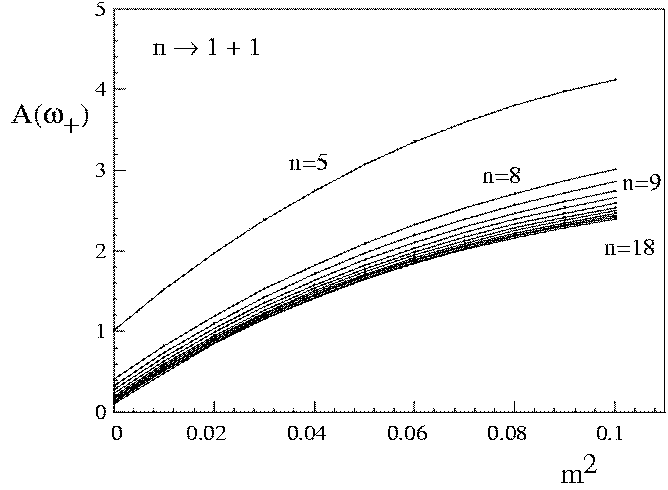}
\renewcommand{\figurename}{Fig.}
\caption{Amplitudes for the decay processes $n\rightarrow 1+1$ 
         ($n = 5, 8, 9, \ldots , 18$) versus the squared fermion mass $m^2$,
         varying from 0 to 0.1.}
\label{fig.1}
\end{minipage}
\end{center}
\end{figure}

\begin{figure}[!ht]
\begin{center}
\begin{minipage}[t]{0.85\textwidth}
\centering
\includegraphics[width=3.2in,angle=-90]{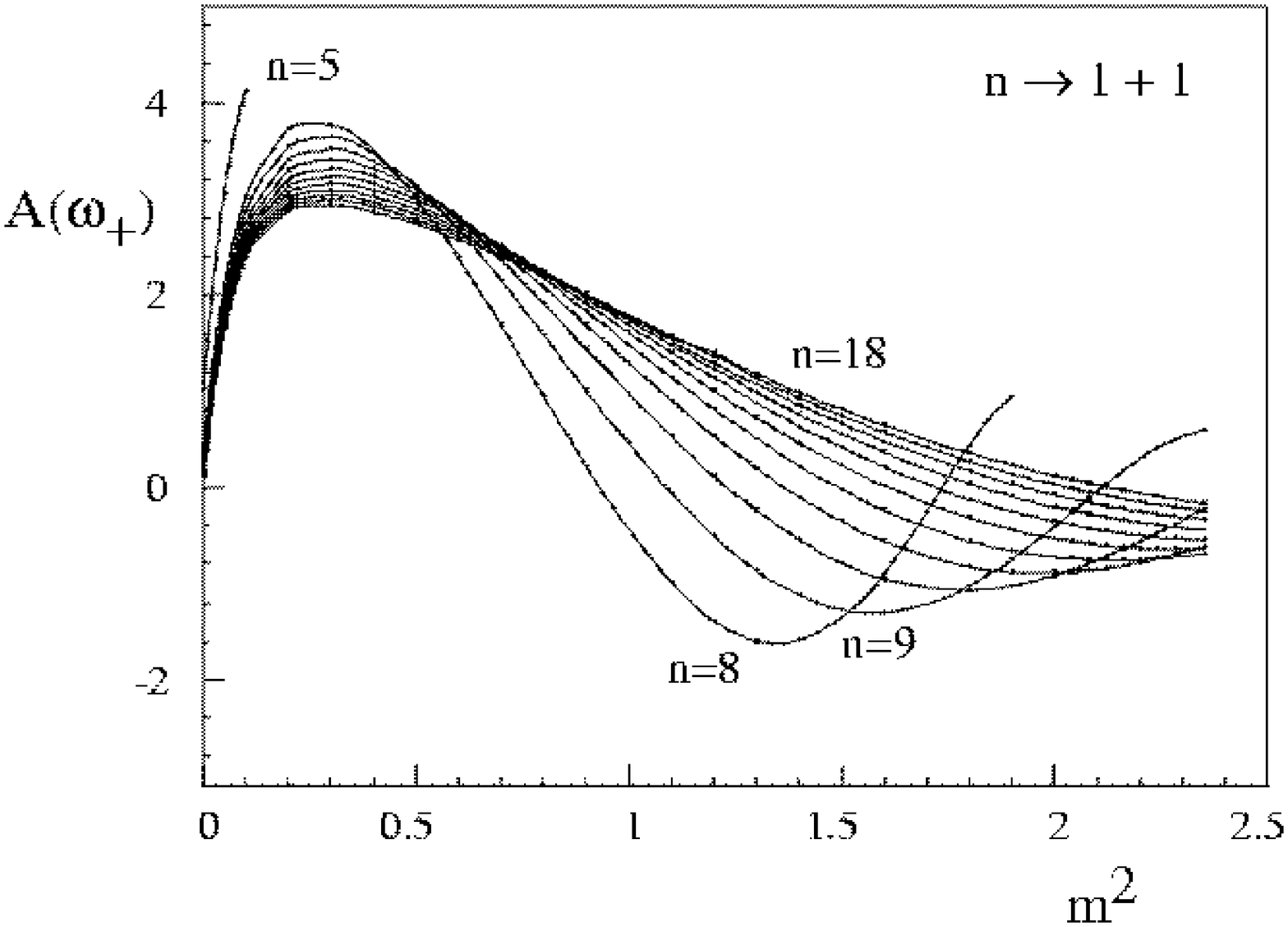}
\renewcommand{\figurename}{Fig.}
\caption{Amplitudes for the decay processes $n\rightarrow 1+1$ 
         ($n = 5, 8, 9, \ldots , 18$) versus the squared fermion mass $m^2$,
         varying from 0 to 2.35. }
\label{fig.2}
\end{minipage}
\end{center}
\end{figure}

Decay amplitudes in the framework of perturbation theory in the inverse
number of colours has been studied by a few authors \cite{fewauth}. The 
essential ingredient is first the solution of  't Hooft equation in order 
to find the bound state wave functions and meson mass-square eigenvalues 
$\mu^2$:
\be
\mu^2 \phi(x) = \frac{\gamma - 1}{x(1-x)} \phi(x) -
       P \int_{0}^{1} dy  \frac{\phi(y)}{(x-y)^2} ~,  \label{hpsi} 
\ee
where $P$ denotes de Cauchy principal value prescription.
 Here $\gamma$ refers to a square mass scale
corresponding to the quark-antiquark pair with equal masses $m^2$,
where the massless limit is obtained for $\gamma =0$ 
\cite{ thooft2}.

\begin{figure}[!ht]
\begin{center}
\includegraphics[width=3.2in,angle=-90]{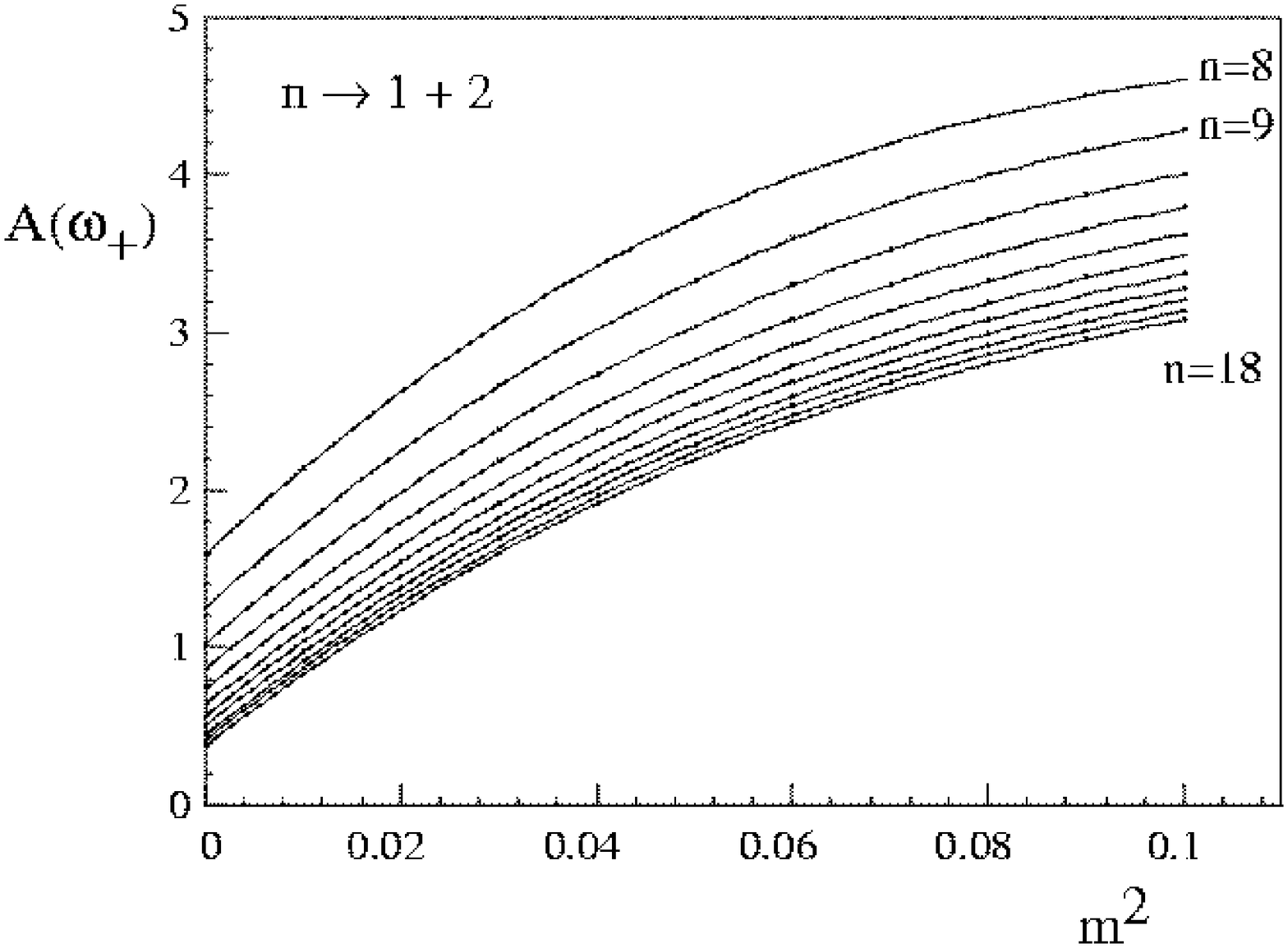}
\renewcommand{\figurename}{Fig.}
\caption{As in figure 1, for the decay processes $n\rightarrow 1+2$
         ($n = 8, 9, \ldots , 18$).}
\label{fig.3}
\end{center}
\end{figure}

\begin{figure}[!ht]
\begin{center}
\includegraphics[width=3.2in,angle=-90]{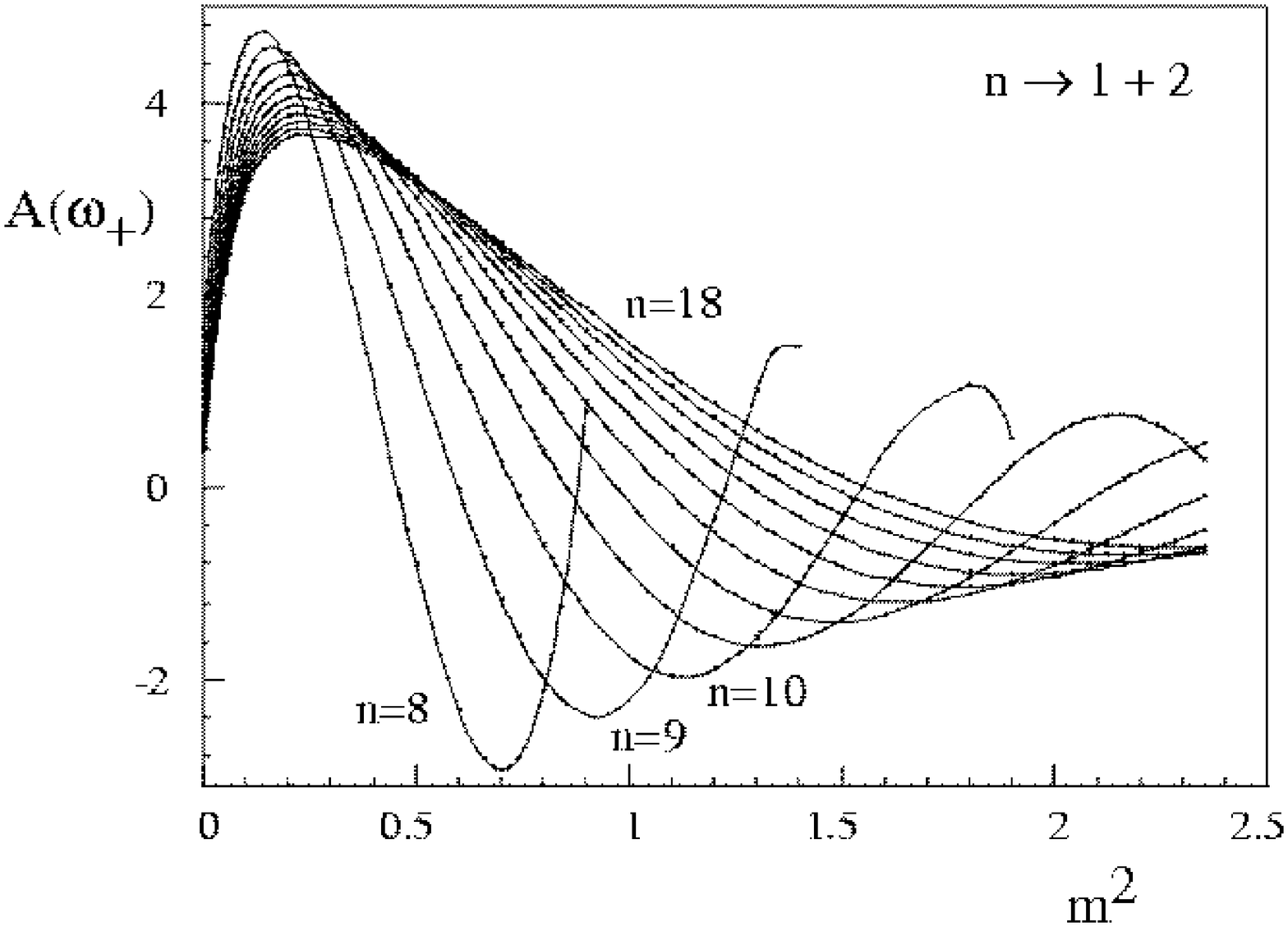}
\renewcommand{\figurename}{Fig.}
\caption{As in figure 2, for the decay processes $n\rightarrow 1+2$
         ($n = 8, 9, \ldots , 18$).}
\label{fig.4}
\end{center}
\end{figure}

\begin{figure}[!ht]
\begin{center}
\includegraphics[width=3.2in,angle=-90]{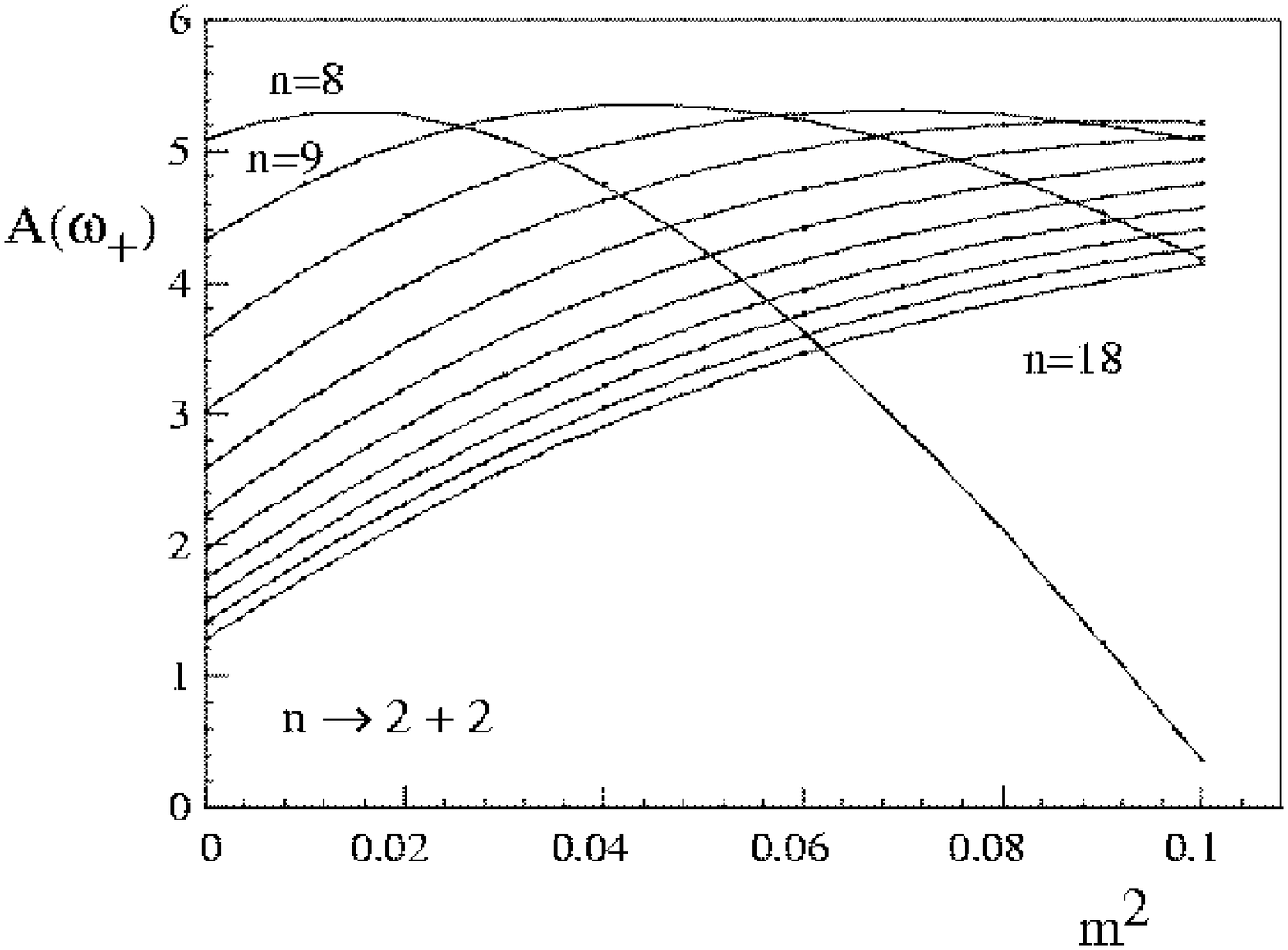}
\renewcommand{\figurename}{Fig.}
\caption{As in fig. 1,  for the decay processes $n\rightarrow 2+2$
         ($n = 8, 9, \ldots , 18$).}
\label{fig.5}
\end{center}
\end{figure}

\begin{figure}[!ht]
\begin{center}
\includegraphics[width=3.2in,angle=-90]{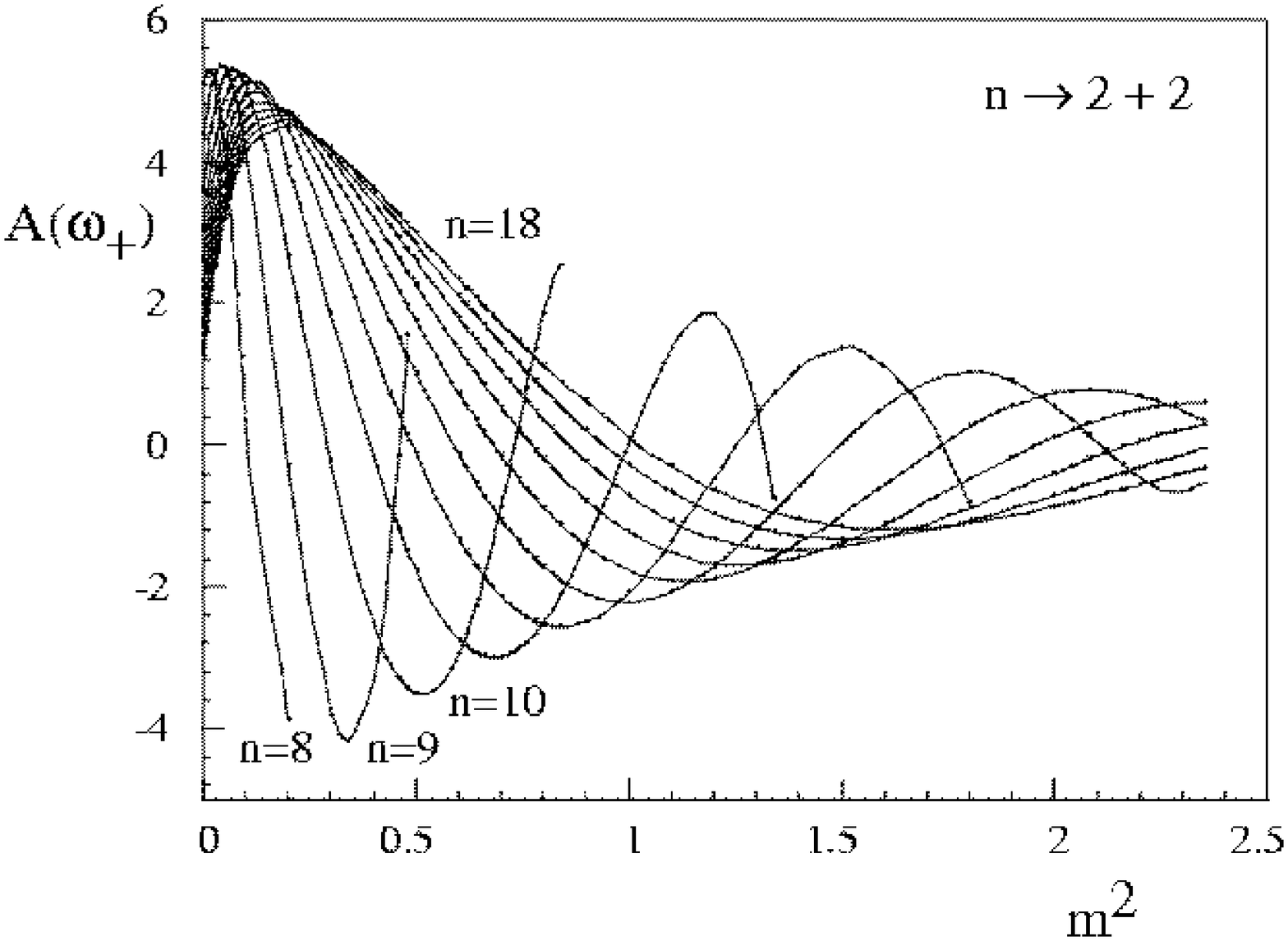}
\renewcommand{\figurename}{Fig.}
\caption{As in figure 2,  for the decay processes $n\rightarrow 2+2$
         ($n = 8, 9, \ldots , 18$).}
\label{fig.6}
\end{center}
\end{figure}

This integro differential equation cannot be solved by analytic methods.
However, using a sensible numerical method presented in  \cite{ks}
one can arrive at quite reasonable results for the eigenvalues and 
wave functions.
  This method makes use of a spline procedure to work out the exact 
wave functions on an adaptive grid. 
  The numerical representation of the eigenfunctions are used to evaluate 
the decay amplitudes from an initial meson state $n$ into two
final states $n_1$ and $n_2$.
  In the leading and next to leading order in the $1/N$ expansion they are
given by \cite{ccg,bd}
\bear
\langle n\vert n_1, n_2\rangle & = & 
   (1-C) \frac{1}{1-w} \int_{0}^{w} dx \phi_{n}(x)\phi_{n_1}(\frac{x}{w})
          \Phi_{n_2}(\frac{x-w}{1-w})       \nonumber \\
& & - (1-C) \frac{1}{w} \int_{w}^{1} dx \phi_{n}(x)\phi_{n_2}(\frac{x-w}{1-w})
          \Phi_{n_1}(\frac{x}{w})           \nonumber \\
& & + \frac{1}{N} (1-C) \frac{f_{n_2}}{1-w} \int_{0}^{w} dx 
        \phi_{n}(x)\phi_{n_1}(\frac{x}{w}) , \label{amplitude} 
\eear
where $C$ denotes de interchange of final states, 
\be
\Phi_{n}(x) = \int_{0}^{1} dy \frac{\phi_{n}(y)}{(x-y)^2}
\ee
 and  $w$ is a kinematic parameter, whose on-shell values $w_{+}$ and
$w_{-}$ correspond to the right moving and left moving of the final 
state $n_1$. 

As observed in \cite{abdmoha} the higher-order corrections for the amplitude
are always multiplied by the factor
\be
f_{n_2} = \int_{0}^{1} dx \phi_{n_2}(x) ,
\ee
where $\phi_{n_2}(x)$ is the 't Hooft's eigenfunction corresponding to the 
decaying state $n_2$, which vanishes for massless fermions (for $n_2\not =0$).

With the knowledge of the eigenfunctions the numerical calculation of these 
integrals is straightforward and we obtain the amplitudes in function of
the outgoing momenta $w$.
From here on we report the behaviour of these amplitudes for the 
on-shell parameter $w_+$,
\be
{\rm A}(w_+)  \equiv \langle n\vert n_1, n_2\rangle_{w=w_+} ~.
\ee 
For this end we had to use an interpolation and extrapolation 
numerical algorithm \cite{press}
to obtain a precise value of A($w_+$) from amplitudes evaluated at a 
discrete set of values for $w$.

First we compute the decay amplitude of a 
level-$n$ state into two level-$1$ states. Notice that level-$0$ state 
decouples for zero fermion mass (see \cite{abdmoha}) and at large $n$ we also 
expect these decay amplitudes to approach zero \cite{abdmoha,schmidt}. As 
argued in \cite{abdmoha} we expect a small anomalous amplitude for vanishing 
mass, and a nonvanishing result for the massive fermion case. Moreover,
there might be a new physical description at some nonvanishing value
of the fermion mass as compared to the (dimensionfull) coupling constant,
namely $m^2\sim \frac{e^2}{\pi}$, since 't Hooft's solution present a 
quark propagator for $m^2< \frac{e^2}{\pi}$. 
 In Fig. 1 we show the results obtained for quark-antiquark pairs with
masses $m^2$ ranging from 0 to 0.1, while in Fig. 2 we extend this range
up to 2.35.
 Interestingly enough are the several zeroes of the amplitudes
for a fermion mass comparable to the coupling constant $e$, but they do 
not coincide for different decays. 
 This behaviour motivated us to continue exploring the decay amplitudes 
involving high excited levels $n$ $(n = 8, 9, ..., 18)$ into the next low-lying
levels $n_1 = 1, n_2 = 2$ (Fig. 3 and 4) and, $n_1 = 2, n_2 = 2$ 
(Fig. 5 and 6).
We observe that the feature described above is reproduced in figures
4 e 6.
 Whether these zeroes signalize any new physics 
hidden by anomalies is an issue about which we can unfortunately only 
speculate. Furthermore, as the bosonic state level increases (resp. 
decreases), it becomes stable at higher (resp. lower) mass values due to
the lack of possibility in terms of kinematical variables, to fulfill
the energy momentum conditions. 
This can be seen as interrupted lines in figures 2, 4 and 6.
 It is a question under 
investigation whether we obtain further stability conditions while correcting  
the mesonic wavefunctions and spectrum in order to cope with the screening
mechanism, and not using a confinement potential between quarks and 
antiquarks, as implicit in 't Hooft's solution.

It is worth to mention the decays $n\rightarrow 2+2$, namely an arbitrary 
state $n$ into two 
mesons at the second excited state present these issues in 
a more enhanced way. Among the studied states, the first  to decay 
according to this mode is the eighth excited
state. The smallness of the decay amplitudes for  $m\sim\frac{e}{\sqrt{\pi}}$
can also be verified and, for not very big values of $n$,
it is clearly seen how the sequence of forbidden decays due to the violation
of the kinematic condition appears as the mass increases.

\section{Conclusions}

After more than two decades of study, two-dimensional QCD largely
remains a challenge for the complete understading of gauge
theories. Nevertheless very deep concepts emerged and a plethora of
information concerning its dynamical understanding could be gathered.

't Hooft's solution taught us how to perform the $1/N$ expansion of the
theory, but also its limitations. From that solution emerged a
massless gauge field, contrary to Schwinger model inputations, and a
spectrum compatible with confinement of the fundamental degrees of
freedom, namely the quarks. The $1/N$  expansion has  been further
studied, and here in particular we  derived the numerical results of
the  decay amplitudes.

It has been revealed that the structure of these amplitudes is
very simple, and in most of the cases rather small, especially for the
massless case, thus compatible with the speculations set forward in
Ref. \cite{abdmoha}. Indeed, we see that in figures 1 through 4 all amplitudes
are small for zero mass, but non vanishing, raising the suspicion of a
higher conservation law, but with the presence of a small quantum
anomaly, effective for the decay of low levels of the meson mass. Therefore,
it turns out to be very important to check whether the 't Hooft hierarchy
is compatible with the screening rather than the confining potential,
in which case the infinite number of states are substituted by a finite
number of mesons. It is to be checked in such a case whether decays
might be further supressed.

Thus, the detailed confirmation of the Regge behaviour and higher states 
in the framework of the $1/N$ expansion,  which is seemingly incompatible 
with the screening potential obtained by several independent authors 
\cite{gross-kle-etc,amz,frishman,frish-sonn,rfm,arfs} requires a
better understanding of the spectrum, and leads to the suspicion that
it contains a finite number of mesonic states. Moreover, it has been
recently pointed out that solitons do not exist in the model,
strenghening the conclusions drawn in \cite{abdmoha} that the symmetries
suspected in \cite{aaijmpa} where anomalously realized. Following this vein, we
confirm that it is essencial to obtain an alternative means of checking the
spectrum of the theory and comparing to 't Hooft's result.
In case we arrive at a confirmation of the latter, we need to explain
the dual description of screening, as obtained from computation of
Wilson loops of semiclassical potential and confinement from the Regge
behaviour of the mesonic states.

From the results of \cite{cr} we also conclude that physical states of the
theory can be constructed out of bilinears. This points to a
confirmation of 't Hooft's spectrum. On the other hand, as we have
seen in the introduction this depends on the rather unclear structure
of the non local constraints we have obtained, which do not inspire
confidence due to the lack of mathematically sound theorems concerning
them.

Further analysis of the diagrams such as figures 5 and 6 point also to
a simplification of the amplitudes for large values of the mass, and 
apparently no transition or anomalous behaviour for the fermion mass
comparable to the coupling constant. At that point, a transition
between weak and strong coupling has been long suspected\cite{aar}, as 
signalized by the existence of a tachyon pole in the (formal) quark propagator.
This points once more to an urgent necessity of further analysis of the
large $N$ limit of two dimensional QCD. A point missed in the large $N$ 
analysis is the fact that for finite $N$ the gauge boson has a mass
generated by the Higgs mechanism, analogous to the Schwinger model
case, and easily derived from the pseudo divergent of the Maxwell
equation with use of the anomaly equation\cite{aar}, but the ensuing mass
vanishes for large $N$.

Although conclusions are still insufficient to have a complete physical
description of the physics of the model, especially in view of the efforts
spent in the problem, progress have been achieved, and in fact the
results already spilt off to the three-dimensional case \cite{abdban},
where the screening mechanism agains prevails, leading to the
suspicion that it is physically more appealing to the theory to form 
kinkstates to take care of the long range force as proposed by \cite{ell-et-al}
rather than to leave for the naked gauge field to pull the quarks
together by means of a long range force. If this kind of mechanism
works also in the four dimensional case the physical consequences will
be far reaching, especially concerning the baryonic and mesonic
spectrum.

{\bf Acknowledgements}: this work was partially supported by Conselho Nacional
de Desenvolvimento Cient\'\i fico e Tecnol\'ogico, CNPq, and FAPESP, Brazil.



\begin{thebibliography}{99} 
\frenchspacing
\bibitem{schw}J. Schwinger, {\it  Phys. Rev.} {\bf 128} (1962) 2425,
{\it Phys. Rev. Lett. }{\bf 3} (1959) 296;
J. Lowenstein and  J. A. Swieca, {\it Annals of Phys. }
{\bf 68} (1971) 172.
\bibitem{aar}E. Abdalla, M.C.B. Abdalla and K.D. Rothe, {\it Non-perturbative
methods in two-dimensional quantum field theory}, World Scientific 1991.
\bibitem{50}A. M. Polyakov and P. B. Wiegmann, {\it Phys. Lett.}  {\bf 131B}
(1983) 121;{\bf 141B} (1984) 223.
\bibitem{wittencmp}E. Witten, {\it Commun. Math. Phys. }{\bf 92} (1984) 455.
\bibitem{thooft}G. 't Hooft,  {\it Nucl. Phys. } {\bf B72} (1974) 461.
\bibitem{thooft2}G. 't Hooft,  {\it Nucl. Phys. }  {\bf B75} (1974) 461.
\bibitem{review}E. Abdalla and M.C.B. Abdalla, {\it Phys. Rep.} 
{\bf 265} (1996) 253.
\bibitem{gross-kle-etc}D. Gross, I. Klebanov and Smilga, {\it Nucl.
Phys.} {\bf B461 } (1996) 109.
\bibitem{amz}E. Abdalla, R. Mohayaee and A. Zadra, {\it Int. J. Mod.
Phys }{\bf A12} (1997) 4539-4557, hepth/9604063.
\bibitem{cr}D. C. Cabra and K. D. Rothe, {\it Phys. Rev. } {\bf D55} (1997)
2240-2246, hep-th/9608155.
\bibitem{crs}D. C. Cabra, K. D. Rothe and F. A. Schaposnik,
{\it Int. J. Mod. Phys. } {\bf A11} (1996) 3379-3391, hep-th/9507043.
\bibitem{aaijmpa}E. Abdalla and M. C. B. Abdalla, {\it Int. J. Mod. Phys.}
{\bf A10} (1995) 1611.
\bibitem{gepner}D. Gepner, {\it Nucl. Phys. } {\bf B252} (1985) 481.
\bibitem{ar-prd}E. Abdalla and K. D. Rothe, {\it Phys. Rev.} {\bf D36} 
(1987) 3190.
\bibitem{dmw}A. Dhar, G. Mandal, S. R. Wadia, 
{\it Phys. Lett.} {\bf B329} (1994) 15-26, hep-th/9403050.
\bibitem{frishman}A. Armoni, Y. Frishman, J. Sonnenschein and U. Trittmann,
hep-th/9805155.
\bibitem{abdmoha}E. Abdalla and R. Mohayaee, {\it Phys. Rev.}
{\bf D57} (1998) 3777-3785, hep-th/9610059.
\bibitem{frish-sonn}Y. Frishman and J. Sonnenschein,  {\it Nucl. Phys. }
{\bf B496} (1997) 285.
\bibitem{fewauth}Benjamin Grinstein and Paul F. Mende, 
{\it Phys. Rev. Lett. } {\bf 69} (1992) 1018-1021,    [hep-ph 9204206];
R.L. Jaffe and Paul F. Mende,  {\it Nucl.Phys.} {\bf B369}
(1992) 189-218.
\bibitem{ks} W. Krauth and M. Staudacher, Phys. Lett. {\bf B388} (1996) 808.
\bibitem{ccg} C.G. Callan, N. Coote and D.J. Gross, Phys. Rev. {\bf D13} 
(1976) 1649.  
\bibitem{bd}  J.C.F. Barbon and K. Demeterfi, Nucl. Phys. {\bf B434} (1995) 
109. 
\bibitem{press} W.H. Press et al., {\it Numerical Recipes} 
(Cambridge University Press, 1989). 
\bibitem{schmidt}R.C. Brower, J. Ellis, M.G. Schmidt and J.H. Weis,
{\it Nucl. Phys.} {\bf B128} (1977) 131 and 175.
\bibitem{rfm}R. Mohayaee, The phases of two-dimensional QED and QCD
hep-th/9705243,  Caribbean Meeting, 1997.
\bibitem{arfs} A. Armoni, Y. Frishman, J. Sonnenschein, hep-th/9807022.
\bibitem{abdban}E. Abdalla and R. Banerjee,
{\it Phys. Rev. Lett.} {\bf  80} (1998) 238-240, hep-th/9704176.
\bibitem{ell-et-al}J. Ellis, Y. Frishman, A. Hanany, M.
Karliner, {\it Nucl. Phys.} {\bf B382} (1992) 189-212,  hep-th/9204212.
\end{thebibliography}
\end{document}